\documentclass[apj]{emulateapj}
\usepackage{bm}
\usepackage{graphicx}
\usepackage{epsf}
\usepackage{graphics}
\usepackage{amsmath}

\def\mpc{\,h^{-1}{\rm Mpc}}
\def\kpc{\,h^{-1}{\rm kpc}}

\def\msun{\,h^{-1}{\rm M}_\odot}
\def\bx{{\boldsymbol x}}
\def\bt{{\boldsymbol t}}
\def\bs{{\boldsymbol s}}
\def\cluster{{\tt cluster$\;$}}
\def\sheet{{\tt sheet$\;$}}
\def\filament{{\tt filament$\;$}}
\def\void{{\tt void$\;$}}

\usepackage{color}

\makeatletter

\newcommand{\Rmnum}[1]{\expandafter\@slowromancap\romannumeral #1@}
\makeatother

\def\cluster {{ \tt cluster$\;$}}
\def\sheet {{ \tt sheet$\;$}}
\def\filament {{ \tt filament$\;$}}
\def\void {{ \tt void$\;$}}

\shorttitle{galaxies in the filaments}
\shortauthors{Zhang et al.}

\begin{document}


\title{Spin alignments of spiral galaxies within the large-scale structure
from SDSS DR7}

\author{Youcai Zhang\altaffilmark{1}, Xiaohu Yang\altaffilmark{1,2}, Huiyuan
  Wang\altaffilmark{3,4}, Lei Wang\altaffilmark{5}, Wentao Luo\altaffilmark{1},
  H. J. Mo\altaffilmark{4}, and Frank C. van den Bosch\altaffilmark{6}}

\altaffiltext{1}{Key Laboratory for Research in Galaxies and Cosmology,
  Shanghai Astronomical Observatory; Nandan Road 80, Shanghai 200030,
  China; yczhang@shao.ac.cn}

\altaffiltext{2}{Center for Astronomy and Astrophysics, Shanghai Jiao
Tong University, Shanghai 200240, China; xyang@sjtu.edu.cn}

\altaffiltext{3}{Key Laboratory for Research in Galaxies and Cosmology,
University of Science and Technology of China, Hefei, Anhui 230026, China}

\altaffiltext{4}{Department of Astronomy, University of Massachusetts,
Amherst MA 01003-9305, USA}

\altaffiltext{5}{Purple Mountain Observatory, the Partner Group of MPI f\"{u}r
  Astronomie, 2 West Beijing Road, Nanjing 210008, China}

\altaffiltext{6}{Department of Astronomy, Yale University, P.O. Box 208101,
  New Haven, CT 06520-8101, USA}

\begin{abstract}
  Using a sample of spiral galaxies selected from the Sloan Digital
  Sky Survey Data Release 7 (SDSS DR7) and Galaxy Zoo 2 (GZ2), we
  investigate the alignment of spin axes of spiral galaxies with their
  surrounding large scale structure, which is characterized by the
  large-scale tidal field reconstructed from the data using galaxy
  groups above a certain mass threshold. We find that the spin axes
  only have weak tendency to be aligned with (or perpendicular to) the
  intermediate (or minor) axis of the local tidal tensor. The signal
  is the strongest in a \cluster environment where all the three
  eigenvalues of the local tidal tensor are positive.  Compared to the
  alignments between halo spins and local tidal field obtained in
  N-body simulations, the above observational results are in best
  agreement with those for the spins of inner regions of halos,
  suggesting that the disk material traces the angular momentum of
  dark matter halos in the inner regions.
\end{abstract}


\section{Introduction}\label{sec_intro}
The galaxy distribution in large spectroscopic surveys, such as Sloan
Digital Sky Survey \citep[SDSS;][]{York2000}, have revealed a complex
hierarchical network of structure, called the cosmic web, composed of
clusters, filaments, sheets, and voids \citep[e.g.][]{Bond1996}. In
the current paradigm of galaxy formation, these structures arise from
the linear growth of Gaussian density fluctuations in a nearly
homogeneous early universe. Dark matter tends to flow out of the
voids, accretes onto the sheets, collapses to the filaments, and
finally accumulates onto the clusters at the intersections of the
filaments. Due to the accretion history of the Universe, the angular
momentum of dark matter halos and galaxies are generated and affected
by their large-scale environment.

According to the tidal torque theory (TTT), protogalaxies acquire
their angular momentum by the asymmetric interaction between their
inertia tensor and the local tidal field \citep{Peebles1969, Doro1970,
  White1984, Hoff1986, Por2002a, Por2002b}.  The galaxy and halo spin
directions are expected to be correlated with the principal axes of
the local tidal tensor.  Over the past decades, both numerical
simulations and real redshift surveys have revealed such correlation
between halos/galaxies and their local large-scale structure
\citep{Fal2002, Cue2008, Paz2008, Zhang2009, WH2011, Cod2012,
  Trow2013, Zhang2013, Tempel2013a, Tempel2013b, Cen2014, Dubo2014}.

Using numerical simulations, \citet{WH2011} found that halo spin
vectors are preferentially aligned with the intermediate axes of the
tidal field, which is consistent with the theoretical prediction using
the linear tidal torque theory. Similar trends are also claimed from
observational data. Based on the real-space tidal field reconstructed
from the 2MASS redshift survey, \citet{Lee2007} claimed that there is
a clear intrinsic alignment between the galaxy spins and the
intermediate principal axes of the tidal tensor.

For halos in filaments from N-body simulations, recent studies
reported a mass-dependence alignment signal between halo spins and
filaments.  The spins of low-mass halos tend to be parallel to
filaments, while high-mass halos have orthogonal alignment
\citep{Arag2007, Hahn2007a, Cod2012, Trow2013}.  For halos in sheets,
the spin vectors tend to lie in the plane of the sheets, independent of
halo mass \citep{Arag2007, Hahn2007a, Hahn2007b, Zhang2009, Lib2012}.

Observationally, the picture is not as clear because of limited
statistics and because of the difficulty of properly defining cosmic
filaments from observational data.  Several studies have endeavored to
solve this problem but give partly contradictory results
\citep{Tru2006, Lee2007, Paz2008, Slo2009, Cera2010, Jone2010,
  Vare2012, Tempel2013a, Tempel2013b}.
For instance, while \citet{Tru2006, Lee2007, Paz2008} found alignment
signals between the spin vectors of galaxies and the surrounding
large-scale structures (e.g. filament and sheet),
\citet{Cera2010, Slo2009} found no statistical evidence for departure
from random orientations.  Using 2dFGRS and SDSS, \citet{Tru2006}
found that spiral galaxies located on the shells of cosmic voids have
their spins aligned preferentially in the void surface, whereas,
\citet{Vare2012}, who used the SDSS DR7 and Galaxy Zoo samples, found
that the spins of spiral galaxies around voids tend to be
perpendicular to the void surface.  Based on SDSS DR7, a recent study
by \citet{Tempel2013a, Tempel2013b} showed that there is a different
alignment between spiral and elliptical galaxies, arguing that the spin
of spiral galaxies tends to lie parallel to their host filaments,
whereas the spin of elliptical galaxies are aligned preferentially
perpendicular to their filaments.

In short, observationally the spin alignment with the surrounding
large-scale structure is still somewhat contradictory and
inconclusive. In our previous paper \citep{Zhang2013}, we examined the
alignment between the major axes of galaxies and their surrounding
large-scale structure. We found that red galaxies are very
significantly aligned with their filaments, while blue galaxies are
only marginally aligned with respect to their filaments.  In this
paper, we investigate the spin vectors of spiral galaxies with respect
to the surrounding large-scale environments.  Here again we use the
galaxy group catalog constructed from the SDSS DR7
\citep{Aba2009}. The cosmic tidal field we used is directly
reconstructed from the group catalog using the method described in
\citet{WH2012}. Based on the signs of the eigenvalues of this tidal
field, we classify the environments of groups into four categories:
clusters, filaments, sheets, and voids. The morphologies of galaxies
used here are those obtained from the Galaxy Zoo 2 catalog
\citep[GZ2;][]{Will2013}.  We measure the spin vectors of spiral
galaxies with respect to the principal axes of the local tidal tensor
and study how these alignments depend on galaxy properties.

The paper is organized as follows. In Section \ref{sec_data}, we
present the observational data and the methodology used to measure the
various alignment signals. The alignments of spiral galaxies with
respect to cosmic web are present in Section \ref{sec_result}.  We
compare the observational results with dark matter halos from N-body
simulations in Section \ref{sec_sim}. Finally we summarize our results
in Section \ref{sec_summary}.

\section{Data and methodology}
\label{sec_data}

In what follows we describe the observational data we adopt to
characterize the cosmic web and measure the alignment between the spin
axes of galaxies relative to the large-scale structures.

\subsection{The SDSS samples}

The galaxy sample we use here is the New York University
Value-Added Galaxy Catalog\footnote{http://sdss.physics.nyu.edu/vagc/}
\citep[NYU-VAGC;][]{Bla2005, Pad2008,Ade2008}, which is based on the
multi-band imaging and spectroscopic survey SDSS DR7
\citep{Aba2009}. As the completion of the survey phase SDSS-II, DR7
features an imaging survey in five bands over a continuous
$7,646\deg^2$ region in the Northern Galactic Cap. The continuity over
this large area is essential and critical to reconstruct the
real-space tidal field from the SDSS survey. From the NYU-VAGC, we
select $639,555$ galaxies in the Main Galaxy Sample with an
extinction-corrected apparent magnitude brighter than $r=17.72$, with
redshifts in the range $0.01 \leq z \leq 0.20$ and with redshift
completeness ${\cal C}_z > 0.7$. A small subset galaxies have
redshifts adopted from the Korea Institute from Advanced Study (KIAS)
Value-Added Galaxy Catalog \citep{Choi2007, Choi2010}.

Using the adaptive halo-based group finder developed by
\citet{Yang2005, Yang2007}, we construct a catalog of $472,532$ galaxy
groups (about $404,300$ groups with single galaxies). For each group,
we assign a halo mass according to the ranking of its characteristic
group luminosity, defined as the total luminosity of all group members
with $^{0.1}M_r - 5 \log h \leq -19.5$, where $^{0.1}M_r - 5 \log h$
is the $r$-band absolute magnitude, $K$-corrected and evolution
corrected to $z=0.1$, using the method described by
\citet{Bla2003}. In order to ensure sample completeness, in what
follows we focus our analysis on the volume covering the redshift
range $0.01 \le z \le 0.12$, and within the largest continuous region
in the Northern Galactic Cap. Using the complete galaxy groups with
mass $M_{\rm h} \ge M_{\rm th} = 10^{12}\msun$, \citet{WH2012}
constructed the mass, tidal and velocity (SDSS-MTV) fields in this
volume. We use the method developed by \citet{WH2009} to correct for
the peculiar motion of groups as well as the finger-of-god effect. We
have tested that our alignment signals are insensitive to the boundary
effect.  The smoothing length scale we used to reconstruct the
  tidal field is $R_s = 2.1\mpc$, as this value provides the best
  agreement with the visual classification of the large-scale
  structure \citep{Hahn2007a, Hahn2007b}.  The smoothing scale $R_s$
  is related to the mass $M_s$ contained in the Gaussian filter at
  mean density $\bar\rho$ via $M_s = (2\pi)^{3/2} \bar{\rho} R_s^3$,
  thus a smoothing scale $R_s=2.1\mpc$ corresponds to the mass
  $M_s=10^{13}\msun$.

\subsection{The Cosmic Web}

Following \citet{WH2012}, we calculate the tidal tensor using 
\begin{equation}\label{TidalTensor}
T_{ij}(\bx) = \frac {\partial^2 \phi} {\partial x_i \partial x_j},
\end{equation}
where $i$ and $j$ are indices with values of $1$, $2$, or $3$, and
$\phi$ is the peculiar gravitational potential, which can be
calculated from the distribution of dark matter halos with mass
$M_{\rm h}$ above some threshold value $M_{\rm th}$ through the
Poisson equation. At the position of each group, the tidal tensor is
diagonalized to obtain the eigenvalues $\lambda_1, \lambda_2$ and
$\lambda_3$ ($\lambda_1 \geq \lambda_2 \geq \lambda_3$), as well as
the corresponding eigenvectors $\bt_1$, $\bt_2$, and $\bt_3$.

According to the Zel'dovich theory \citep{Zeld70}, the sign of the
eigenvalues of local tidal tensor can be used to classify the group's
environment into one of four categories \citep{Hahn2007a, Hahn2007b}:
\begin{itemize}
  \item {\cluster}: a point where all three eigenvalues are positive;
  \item {\filament}: a point where $T_{ij}$ has one negative and two
    positive eigenvalues;
  \item {\sheet}: a point where $T_{ij}$ has two negative and one
    positive eigenvalues;
  \item {\void}: a point where all three eigenvalues are negative.
\end{itemize}
The direction of a filament at the location of a \filament group can be
identified with the eigenvector corresponding to the single negative
eigenvalue. Likewise, the normal vector of a sheet for a \sheet group
is the eigenvector corresponding to the single positive eigenvalue of
the tidal tensor.

\subsection{Spin vectors of galaxies}
\begin{figure}
\center
\includegraphics[width=0.48\textwidth]{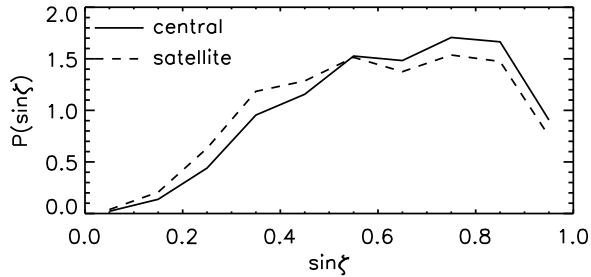}
\caption{The probability distribution of the inclination angles of galaxies.
 The solid line shows the distribution of $74,713$ central galaxies;
 the dashed line shows the distribution for $28,514$ satellite 
 galaxies.}
\label{fig:ia}
\end{figure}

We adopt the morphological classifications of galaxies from the Galaxy
Zoo $2$ (GZ2) catalog
\citep{Will2013}\footnote{http://data.galaxyzoo.org}, which provides
robust classifications into elliptical and spiral for $304,122$
galaxies. In the following analysis, we only use spiral galaxies with
$gz2\_class$ strings beginning with 'S'. This results in a sample of
$120,102$ spirals in our survey volume.  Based on group catalog, the
sample contains $90,761$ centrals and $29,341$ satellite galaxies,
among which $74,713$ centrals and $28,514$ satellites have reliable
halo masses.  To compare with the results from N-body simulations, we
focus on the $74,713$ central galaxies with reliable halo
masses. Based on local tidal fields described in the last subsection,
$12,386$ galaxies ($16.6\%$) are located in \cluster environments,
$49,898$ ($66.8\%$) in \filament, $11,909$ ($15.9\%$) in \sheet, and
$520$ ($0.7\%$) in \void.

In order to obtain a more reliable detection of the spin vectors $\bs$
of galaxies, in this paper we only use a sample of spiral galaxies.
If a spiral galaxy were a thin circular disk, the inclination angle
$\zeta$, between the plane of the disk and the line of sight, can be
determined from the projected minor-to-major axis ratio $b/a$.  Many
studies used a simple model for the finite disk thickness to relate
the apparent axis ratio $b/a$ and an intrinsic flatness parameter $f$
to obtain the inclination angles \footnote{Note that in some analyses,
  the inclination angle $i$ is defined as the angle between the line
  of sight and the normal to the galaxy disk, so that
  $i=90^{\circ}-\zeta$.}, $\zeta$, of galaxies \citep{Hay1984,
  Lee2007, Vare2012}.  Here we follow the same practice and obtain
$\zeta$ through
\begin{equation}\label{IncAng}
\sin^2\zeta = \frac{(b/a)^2 - f^2} {1-f^2}
\end{equation}
for $b/a>f$ and set $\zeta=0$ for $b/a<f$. According to
\citet{Hay1984}, the intrinsic flatness $f$ depends on the
morphological type of galaxies. However, we will use an average
value of $0.14$ \citep{Vare2012}, since the variation of the flatness
$f$ does not affect the calculation of the inclination angle
$\zeta$ significantly. We have repeated our entire calculation 
of the alignment assuming $f=0$, and found that the results 
are very similar to those using $f=0.14$.

Figure~\ref{fig:ia} shows the probability distribution of the
inclination angles of galaxies. We find that the distribution differs
significantly from a uniform distribution. This is mostly likely due
to the fact that spiral galaxies have bulges which distort the
projected axis ratio. If the $b/a$ axis ratio is measured around an
isophote where the bulge contributes significantly, then the result
can be affected. This effect becomes more and more significant for
systems that are close to edge-on.  Because of this, our following
analysis will focus on galaxies with $\sin\zeta>0.5$, which results in
a sample of $54,457$ centrals and $18,976$ satellites.

From a two-dimensional image of a galaxy, projected on the sky, we can
not decide which side of the galaxy is closer to us, which means that
in the determination of $\zeta$ using equation~(\ref{IncAng}), a
single value of $b/a$ corresponds to two values of the inclination
angle: $\pm|\zeta|$. To avoid this problem, many studies constrain
their analysis to edge-on ($\zeta\sim0$) and face-on
($\zeta\sim\pm\pi/2$) galaxies, which usually results in small samples
of about a few hundreds of galaxies \citep[e.g.][]{Tru2006, Slo2009,
  Jone2010}.  In this paper, we use all the spirals selected from GZ2
catalog. Throughout we apply the positive sign of $\zeta$ for all the
spirals.  We have tested and found no difference in the alignment
signal if using the negative sign for $\zeta$. However, the sign
ambiguity is expected to reduce the strength of the alignment signal,
as we quantify in the following section.

Following \citet{Tru2006, Lee2007, Vare2012}, we compute 
the components of the spin vector $\bs$ as
\begin{align}\label{spin}
s_x & = \cos\alpha \cos \delta \sin \zeta \nonumber \\
    & + \cos \zeta (\sin\phi \cos \alpha \sin \delta - \cos \phi \sin \alpha) \nonumber \\
s_y & = \sin \alpha \cos \delta \sin \zeta \nonumber \\
    & + \cos \zeta (\sin\phi \sin \alpha \sin \delta + \cos \phi \cos \alpha) \nonumber \\
s_z & = \sin \delta \sin \zeta - \cos \zeta \sin \phi \cos \delta,
\end{align}
where $\alpha$ and $\delta$ are the right ascension and
declination of a galaxy, respectively, $\zeta$ is the inclination 
angle calculated from equation~(\ref{IncAng}). 
The position angle $\phi$ increases counterclockwisely from 
north to east in the plane of the sky, and is specified by the $25$ 
magnitudes per square arc-second isophote in the $r$-band. 

\subsection{Computation of the alignment}
\begin{figure*}
\begin{center}
\includegraphics[width=0.49\textwidth]{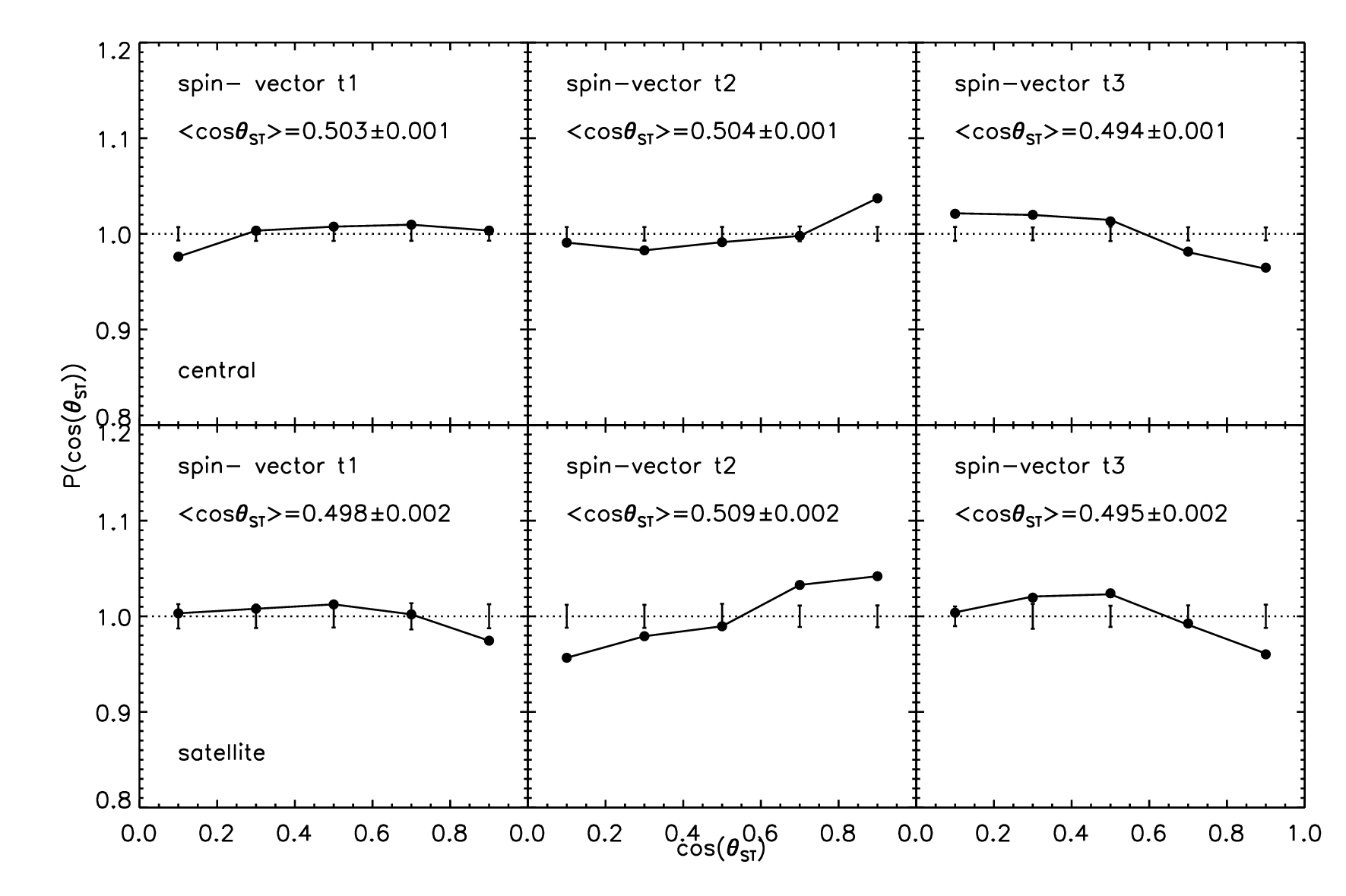}
\includegraphics[width=0.49\textwidth]{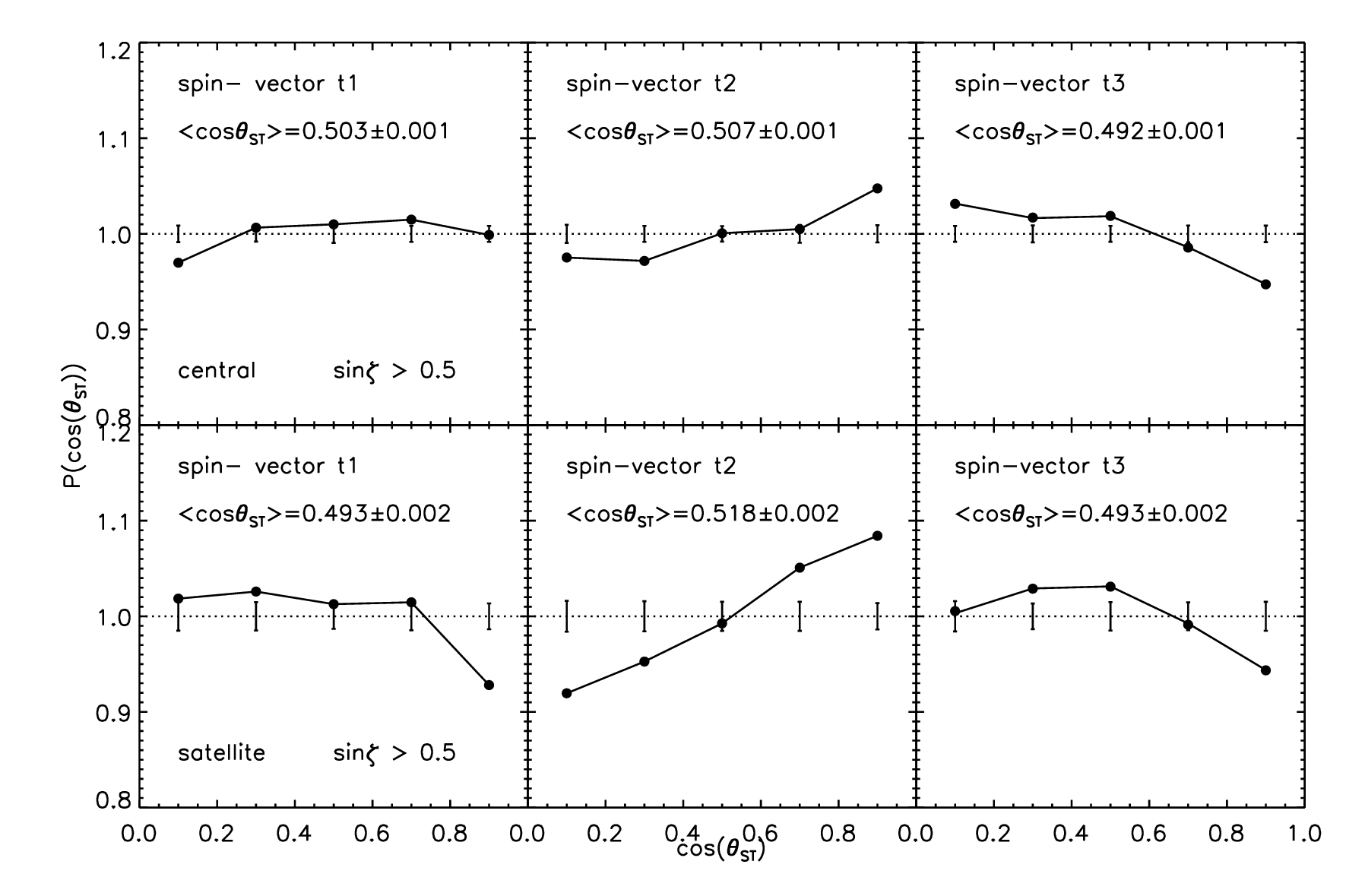}
\end{center}
\caption{Normalized pair count as a function of cosine of the angle
  between the galaxy spin vector and the eigenvectors $\bt_1$ (left),
  $\bt_2$ (middle), and $\bt_3$ (right) of the tidal tensor field for
  spiral galaxies from GZ2 and SDSS. The panel on the left contains
  $74,713$ centrals (upper panels) and $28,514$ satellites (lower
  panels), whereas the panel on the right repeats the measurements for
  galaxies with $\sin\zeta>0.5$.  The eigenvector $\bt_i$ corresponds
  to the eigenvalue $\lambda_i$, where $\lambda_1 \geq \lambda_2 \geq
  \lambda_3$.  The horizontal dotted line corresponds to null
  hypothesis, while the error bars indicate the scatter obtained from
  $100$ realizations in which the spin vectors of the galaxies have
  been randomized. The average value of $\cos\theta_{\rm ST}$ and its
  error (obtained from the $100$ random realizations) are indicated in
  the panels.  }
\label{fig:tf}
\end{figure*}

To estimate the correlation between spin vectors of galaxies and the
large-scale structure, we compute the cosine of the angle,
$\theta_{\rm ST}$, defined as the angle between the spin axis of the
galaxy, $\bs$, and the three principal axes $\bt_1$, $\bt_2$, and
$\bt_3$ of the local tidal tensor.  To quantify the alignments and
their significance, we generate $100$ random samples in which the
orientations of the principal axes of the tidal tensors are kept
fixed, but the spin axes of the galaxies are randomized. The alignment
strength is then expressed by the normalized pair count (e.g. Yang et
al. 2006):
\begin{equation}
P(\cos\theta_{\rm ST}) = \frac {N(\cos\theta_{\rm ST})}
{\langle N_{\rm R}(\cos\theta_{\rm ST}) \rangle},
\end{equation}
where $N(\cos\theta_{\rm ST})$ is the number of pairs for each bin in
$\cos\theta_{\rm ST}$, and $\langle N_{\rm R} (\cos\theta_{\rm ST})
\rangle$ is the average number of such pairs obtained from the $100$
random samples.  We use $\sigma_R(\cos\theta_{\rm ST})/\langle
  N_{\rm R} (\cos\theta_{\rm ST}) \rangle$, where
  $\sigma_R(\cos\theta_{\rm ST})$ is the s.d. of $N_{\rm R}
  (\cos\theta_{\rm ST})$ obtained from the 100 random samples, to
  assess the significance of alignment in each bin.  Since the
significance is quantified with respect to the null hypothesis, we
plot the error bars on top of the $P(\cos\theta_{\rm ST}) = 1$
line. In addition to the normalized pair count, we also calculate the
mean angle $\langle \cos\theta_{\rm ST} \rangle$ and
$\sigma_{\cos\theta_{\rm ST}}$, which is the standard deviation of
${\langle \cos\theta_{\rm ST} \rangle}_{\rm R}$ for the $100$ random
samples.  In this paper, $\cos\theta_{\rm ST}$ is restricted to the
range $[0,1]$. In the absence of any alignment, $P(\cos\theta_{\rm
  ST})=1$ and $\langle \cos\theta_{\rm ST} \rangle = 0.5$.
$\cos\theta_{\rm ST} = 1$ implies that the galaxy spin is parallel to
the vector in question, while $\cos\theta_{\rm ST} = 0$ indicates it
is perpendicular to it.

\section{Results}
\label{sec_result}

\subsection{Alignment between Galaxy Spin and Tidal Tensor}

\begin{figure}
\center
\includegraphics[width=0.50\textwidth]{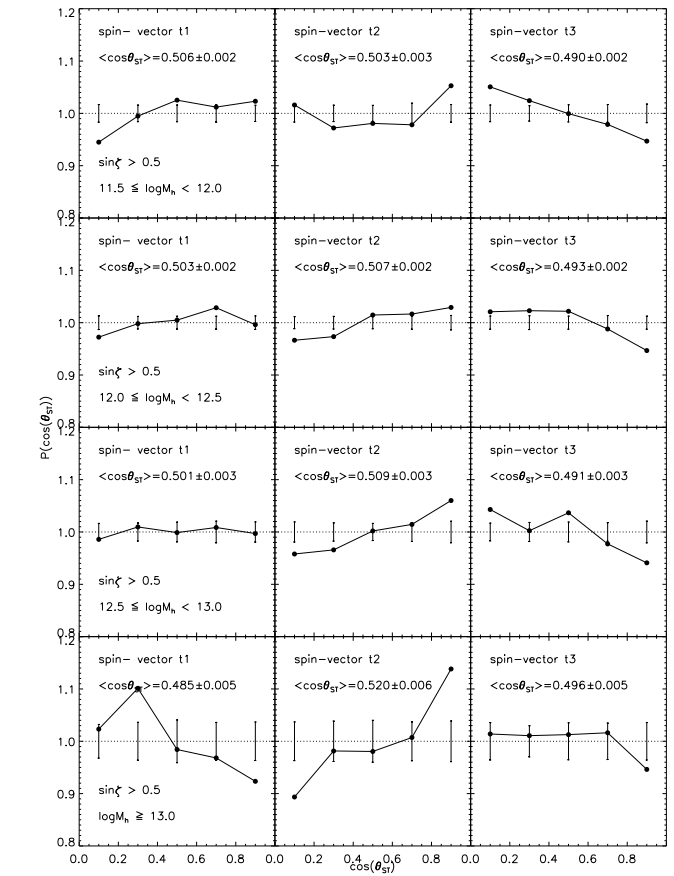}
\caption{Same as Figure~\ref{fig:tf} but for central galaxies 
in halos (groups) of different  masses. 
Here results are shown only for galaxies with $\sin\zeta>0.5$.
}
\label{fig:tf_mass}
\end{figure}

\begin{figure}
\center
\includegraphics[width=0.50\textwidth]{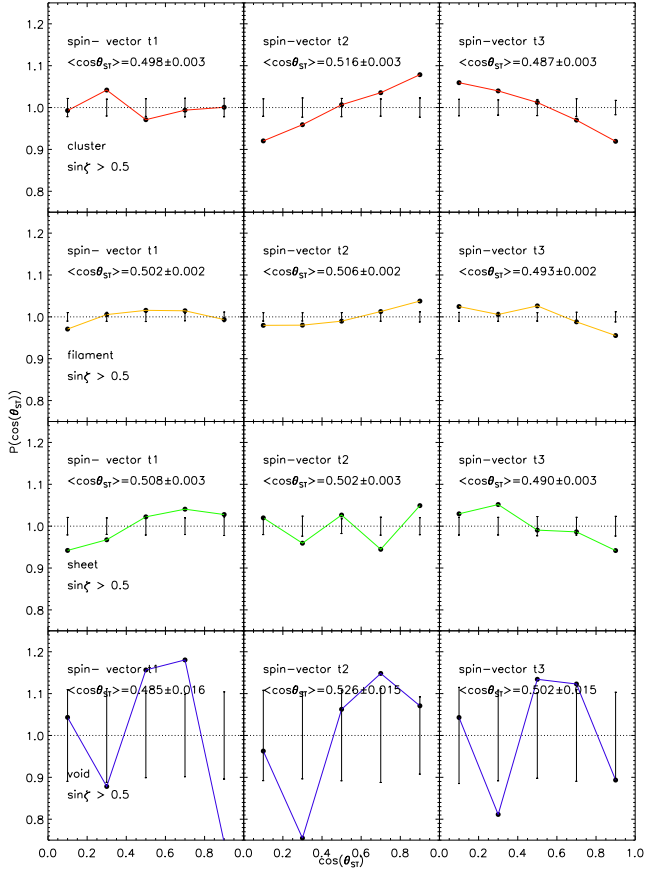}
\caption{Same as Figure~\ref{fig:tf} but for galaxies in clusters (first row), 
filaments (second row), sheets (third row), and voids (fourth row).
Here results are shown only for galaxies with $\sin\zeta>0.5$.
}
\label{fig:tf1}
\end{figure}

Figure~\ref{fig:tf} shows the normalized pair count $P(\cos\theta_{\rm
  ST})$ of the cosine of the angle between the galaxy spin vector and
the eigenvectors $\bt_1$, $\bt_2$, and $\bt_3$ of the local tidal
tensor. In the left panel, the calculation of the alignment is based
on all $74,713$ centrals and $28,511$ satellite spiral galaxies
selected from GZ2.  In the right panel, we repeat the measurements for
galaxies with $\sin\zeta>0.5$.  As expected, galaxies with
$\sin\zeta>0.5$ have stronger alignment signals because of the reduced
uncertainty in the values of $\zeta$.  The average cosines and their
errors are displayed for reference.

Figure~\ref{fig:tf} shows that the galaxy spin vector has a weak
tendency to be aligned with the intermediate axis $\bt_2$, of the
local tidal tensor, which is consistent with the theoretical
prediction using the linear tidal torque theory \citep{Lee2007}.
Similar trends have been also found in simulation data. Based on the
tidal field traced by halos from N-body simulation, \citet{WH2011}
found that the spin vector of halos with masses larger than
$10^{12}\msun$ shows significant alignment with the intermediate axis
of the tidal field. Using a cosmological hydrodynamic simulation of
galaxy formation in a cosmic filament, \citet{Hahn2010} found that the
spin vector has a weak alignment with the intermediate axis of the
tidal field for their medium mass sample at $z=0$ (see their
figure~$11$). We can also see that the galaxy spin tends to be
perpendicular to the minor axis $\bt_3$, which is consistent with the
result shown in the right panel of figure~$2$ in \citet{Lee2007}.  The
alignment with the major axis $\bt_1$ is almost random.

It is interesting to see that the alignment signal for satellites is
as strong as that for centrals. This may suggest that satellites can
preserve their alignment with the large scale tidal field after they
have fallen into their host halos.

Since we have the halo mass for each central galaxy, we can
investigate how the spin-tidal field alignment signal depends on the
halo mass of groups.  To that extent we split the central galaxies
into four sub-samples of halo masses. Figure~\ref{fig:tf_mass} shows
the alignment signals as a function of halo mass.  As one can see, the
alignment signals in middle panels appear to be slightly stronger for
centrals in more massive halos.  It's also interesting to see
  that there is a transition of the alignment signals in the lower
  left panel, although the error bars are very large due to the small
  number galaxies with halo mass larger than $10^{13}\msun$. 

\subsection{Alignment of galaxies in different environments}

Figure~\ref{fig:tf1} shows the alignment between the three
eigenvectors of the local tidal tensor and the spin vector of central
galaxies which are located in cluster (red), filament (orange), sheet
(green), and void (blue) environments. Here we see that for spiral
galaxies in \cluster environments (first row), the spin vectors have a
significant alignment with the intermediate axis of the local tidal
tensor.   The upper right panel indicates that the spin vectors of
  galaxies in \cluster environments have a strong anti-alignment with
  the minor axis of the local tidal tensor. For galaxies in \sheet
and \filament environments, the alignments between spin vectors and
the intermediate eigenvectors are rather weak. Due to the small number
of galaxies in \void environments, we can not obtain any useful
information for these galaxies.

According to the definition of \filament, the right panel in the
second row in Figure~\ref{fig:tf1} can be considered as the normalized
pair count for the cosine of the angle between the spin vector and the
direction of the filament in which the galaxy resides.  There is a
weak tendency that \filament galaxies tend to have their spins
preferentially perpendicular to their filaments, consistent with the
prediction from the tidal torque theory.  However, some of previous
investigations found opposite alignments. Using $69$ edge-on galaxies
($b/a<0.2$) in filaments detected from SDSS DR5, \citet{Jone2010}
found that only $14$ galaxies have their spin axes aligned
perpendicular to their host filaments. This indicates that the spin of
the edge-on galaxies tends to be parallel to their hosted
filaments. However, as we have tested, if one only use edge-on or
face-on galaxies, one needs to have a fairly large sample of filament
directions to get statistically meaningful results.  Otherwise the
signals may not be reliable. For example, if we have only one filament
which lies along the line of sight direction, the edge-on or face-on
galaxies will all have perfect perpendicular or alignment signals by
selection.

Based on a galaxy sample constructed from SDSS DR8,
\citet{Tempel2013b} found that the spin axes of spiral galaxies tend
to align with the host filaments, which is opposite to our alignment
signal. This discrepancy may be due to the different methods used to
identify filamentary structures. In their investigation, filaments are
traced by thin cylinders placed on the galaxy number density field,
while the filaments in our samples are obtained from the continuous
potential field constructed with the use of the method of
\citep{WH2009}.

For spiral galaxies in \sheet environments, the left panel in the
third row in Figure~\ref{fig:tf1} may be considered as the normalized
pair count for the cosine of the angle between the spin and the normal
of the sheet in which the galaxy resides.  The results indicate that
the spin vectors of \sheet galaxies have a very weak tendency to align
along the normal of the sheet.  Using a sample of disk galaxies lying
in the surfaces of voids obtained from SDSS DR7, \citet{Vare2012}
found a significant tendency of galaxies around very large voids to
have their spins aligned with the radial direction of the voids. This
trend is consistent with our results. Using $178$ edge-on and $23$
face-on galaxies from the 2dFGRS and SDSS DR3, \citet{Tru2006} found
that spiral galaxies located on the walls of the largest voids have
spins lying preferentially on the void surface.  The disagreement may
again be due to the small number of galaxies \citet{Tru2006} used.

\section{Comparison with N-body Simulations}
\label{sec_sim}
\begin{figure}
\center
\includegraphics[width=0.5\textwidth]{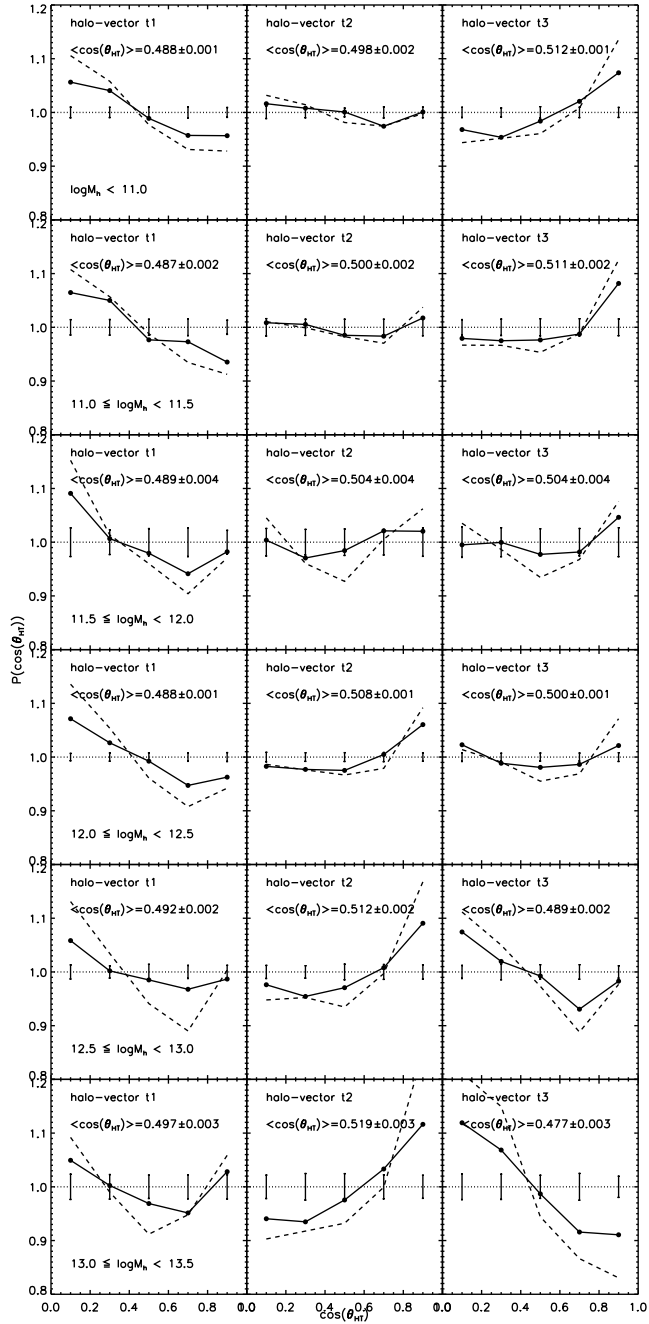}
\caption{Normalized pair count of the cosine of the angle between the
  halo spin vector and the eigenvectors $\bt_1$ (left), $\bt_2$
  (middle), and $\bt_3$ (right) of the tidal tensor field for FOF
  halos.  The dashed and solid line correspond to the original and
  control halo catalog, respectively. In the control sample, the spin
  vectors of $50\%$ halos have been changed (see text for a detailed
  description).  }
\label{fig:FOF}
\end{figure}
\begin{figure}
\center
\includegraphics[width=0.5\textwidth]{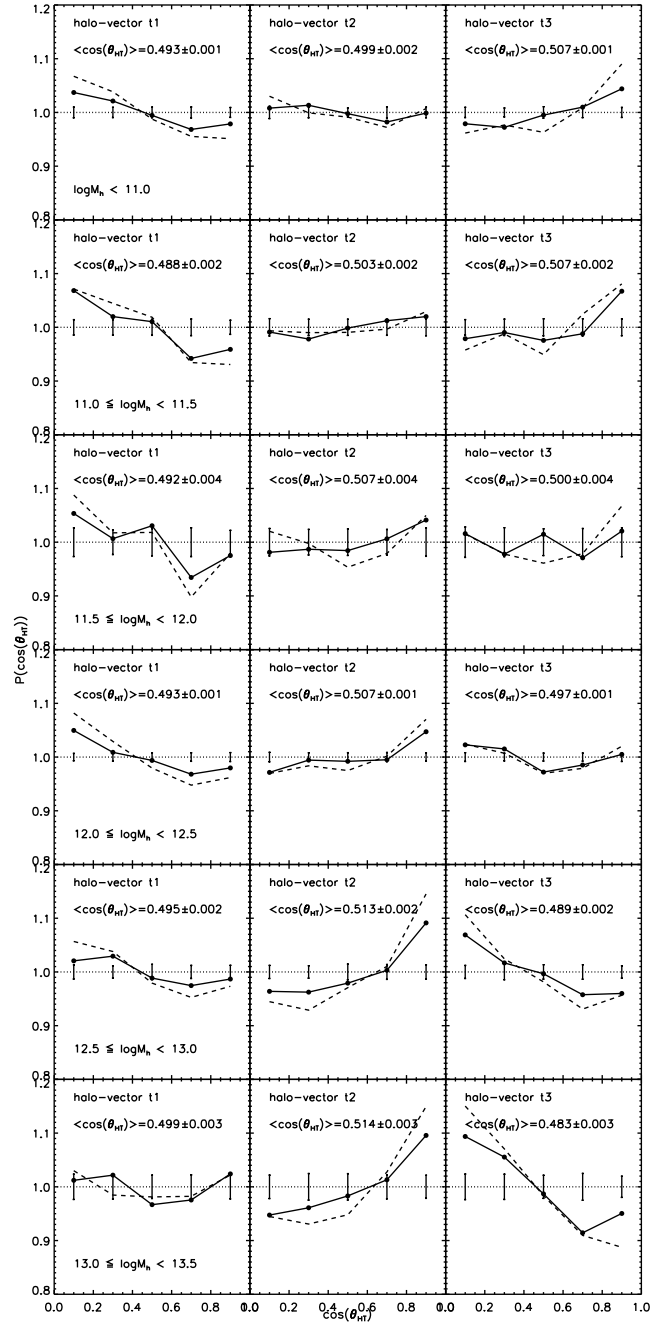}
\caption{The same as Figure~\ref{fig:FOF}, but for SO halos with $\Delta=200$.  }
\label{fig:SO200}
\end{figure}
\begin{figure}
\center
\includegraphics[width=0.5\textwidth]{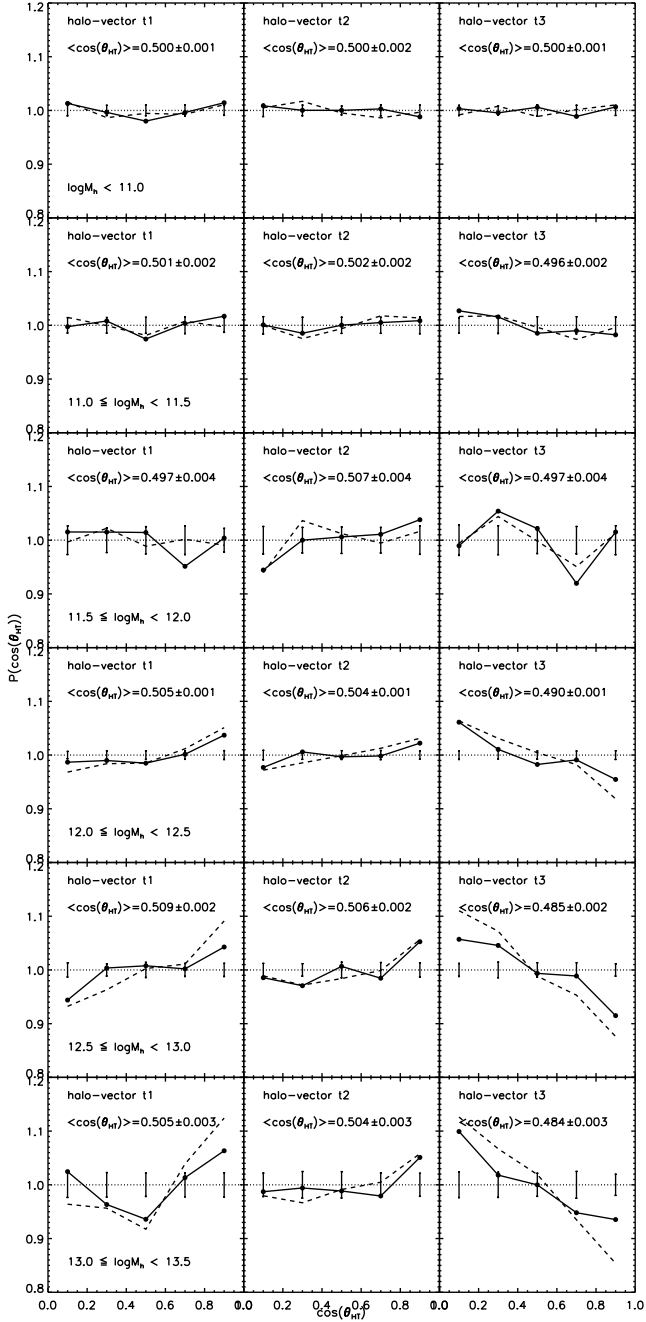}
\caption{The same as Figure~\ref{fig:FOF}, but here spin vectors 
are estimated for inner halos within which the over density  is 
$\Delta=2500$. }
\label{fig:SO2500}
\end{figure}

To facilitate the interpretation of our results, we adopt two
different N-body simulations using the massively parallel TreeSPH 
GADGET-2 code \citep{Springel2005}. The simulations evolved 
$1024^3$ dark matter particles in $100\mpc$ and $300\mpc$ periodic
boxes. The `glass-like' initial conditions \citep{White1996} were
generate using the Zel'dovich approximation at redshift $z=127$.
The simulations adopted the cosmological parameters with 
$\Omega_{\rm m} = 0.258$, $\Omega_{\rm b} = 0.044$, $\Omega_\Lambda 
= 0.742$, $h=0.719$, $n=0.963$ and $\sigma_8= 0.796$. 

For the $100\mpc$ box simulation, the mass resolution and softening lengths 
are, respectively, $6.67\times10^7\msun$ and $3\kpc$. This small 
volume simulation can give better statistics for halos with mass 
less than $10^{12}\msun$. The $300\mpc$ box simulation
(mass resolution of $1.8\times10^9\msun$) is used for computing the
alignment signals for halos with mass larger than $10^{12}\msun$.

Dark matter halos are identified at redshift $z=0$ using the standard
FOF algorithm \citep{Davis1985} with a linking length of $0.2$ times
the mean inter-particle separation.  To obtain a reliable measurement,
only halos with more than $500$ dark matter particles are included for
further analysis, resulting in a catalog including $72,783$ ($103,001$)
halos with mass larger than $3.5\times10^{10} \msun$ 
($9.0\times10^{11} \msun$) for $100\mpc$ ($300\mpc$) simulation.
As comparison, we also use the SO (spherical over-density) 
algorithm to identify halos. Beginning with the most
bound particles from the FOF halos, a sphere is grown around the
particle until the mean interior density is equal to the input
value $\Delta$, where $\Delta$ is the over-density within a sphere
with respect to the critical density $\rho_{\rm crit}$ of the universe.
Here, we adopt $\Delta=200$ and $\Delta=2500$,
representing the entire halo and the inner region, respectively.

To obtain a spin vector of a halo containing $N$ particles, we
compute the total angular momentum $\boldsymbol S$ using
\begin{equation}
\boldsymbol S = m \sum_{i=1}^N \boldsymbol r_i \times \boldsymbol v_i,
\end{equation}
where $m$ is the mass of a dark matter particle, $\boldsymbol r_i$
denotes the position vector of particle $i$ relative to the center of
mass, and $\boldsymbol v_i$ is the velocity relative to the bulk
velocity of the halo.

As mentioned before, the spin vectors of galaxies in observation are
not uniquely defined because we do not know which side of the galaxy is
closer to us. In order to check this projection effect, we construct a
control halo catalog in which we have changed the spin vectors of
$50\%$ halos randomly selected from the original halo catalog. 
The changed spin unit vector ${\boldsymbol s_c}$ is computed by
\begin{equation}
{\boldsymbol s_c} = 2{\boldsymbol p}\cos\theta - {\boldsymbol s}
\end{equation}
where $\cos\theta = \boldsymbol s \cdot \boldsymbol p$, 
$\boldsymbol s$ is the original spin unit vector of a halo, and 
$\boldsymbol p$ is the position unit vector of the halo, where
we place a virtual observer at the corner of the simulation box.
Note that the original and changed spin vector of a halo are 
symmetric with respect to the line-of-sight direction 
${\boldsymbol p}$.

In the simulation, we have classified the cosmic web environment of a
dark matter halo using the Density Field Hessian Matrix Method
\citep{Zhang2009}, similar to the method used above based on the tidal
tensor. Here, in order to compare with the alignment signals detected
in the SDSS data, we compute the probability distribution function
$P(\cos\theta_{\rm HT})$, where $\theta_{\rm HT}$ is the angle between
the halo spin and the eigenvectors of the tidal tensor.

Figure~\ref{fig:FOF} shows the normalized pair count $P(\cos
\theta_{\rm HT})$ of the cosine of the angle between the halo spin and
the eigenvectors of the tidal tensor. The solid line corresponds to
the control sample, in which the spin vectors of $50\%$ halos have
been changed to model the projection effect in observation. The dashed
line denotes the alignment signal from the original halo catalog in the
N-body simulation, which is consistent with the results described in
\citet{Zhang2009}. From Figure~\ref{fig:FOF}, we find that the 
alignment strength for control sample decreases about $50\%$, which 
means the indetermination of the spin vector in observation can 
reduce the alignment strength significantly.

The middle panels in Figure~\ref{fig:FOF} show that the halo spin 
vector is aligned with the intermediate axis of the tidal tensor, 
and the strength of the alignment is stronger for higher mass halos. 
The results plotted in the right panels show that halo spin vector 
is aligned with the minor axis of the tidal field for low mass halos
but tends to be perpendicular to it for massive halos. 
This mass dependence is consistent with recent results 
obtained for halo spin alignments with respect to filaments
\citep{Arag2007, Hahn2007a, Cod2012, Trow2013}. 
As comparison, Figure~\ref{fig:SO200} shows the alignment 
signals for SO halos with the over-density $\Delta=200$. 
The strength of the alignment signals is only slightly weaker 
than that for FOF halos. The observational alignment signals shown in
Figure~\ref{fig:tf} are in general agreement with N-body simulation
results, in that only galaxies in halos with masses larger than 
$10^{12.0}\msun$ show significant alignments with the intermediate 
axis. 
Figure~\ref{fig:SO2500} shows the alignment signals 
for SO halos where spin vectors are calculated only 
by using the inner regions within which the over-density 
$\Delta=2500$. Here we see that the alignment signals 
are weaker than those for the entire halos. In particular
the tendency that spin vectors are perpendicular to the 
major axis seen for the total FOF and SO halos is now 
replaced by a  weak alignment. There is a stronger tendency 
for spin axes of the inner halos to be perpendicular to $\bt_3$ 
for halos with mass larger than $10^{12}\msun$. These 
results match better the observational results for spiral 
galaxies, suggesting that disk material may follow better 
the angular momentum of the halo material in the inner parts. 

\section{SUMMARY}
\label{sec_summary}

Using spiral galaxies selected from the SDSS DR7 and the GZ2, we have
investigated the alignment between the spin of galaxies and their
surrounding large-scale structure, which is characterized by the
local tidal tensor.  Based on the eigenvalues of the tidal tensor at the
location of each galaxy group, we have classified the galaxy
population into four different environments: \cluster, \filament,
\sheet and \void. We have found that the spin axes of central 
spiral galaxies selected from the GZ2 have a weak tendency 
to be aligned with the intermediate direction of the local tidal tensor. 
This is consistent with the theoretical prediction using the linear 
tidal toque theory \citep{Lee2007}.  There is almost no 
significant alignment between the spin axes of spiral galaxies 
and the major axis $\bt_1$ of the local tidal tensor, 
while the spins of spiral galaxies have a weak tendency to be 
perpendicular to the minor axis $\bt_3$ vector (defined as the
direction of the eigenvector of the tidal tensor associated with the
single negative eigenvalue $\lambda_3$).  

The galaxies in \cluster environment show the most prominent
alignment signals. For all spiral galaxies in
\filament environment, the spin axes show a weak 
tendency to be perpendicular to their filaments. 
For galaxies in \sheet environment, there is
a weak tendency for their spins to align with the norm of
the sheet.

Using large N-body simulations, we check the projection effect in
observation by constructing a control halo catalog, in which the spin
vectors of $50\%$ halos have been changed to mimic the fact that 
it is impossible to determine which side of the galaxy is closer 
to us from the observed axis ratio.  
This indetermination can reduce the alignment strength significantly.
The simulations also show that the spin vectors of the 
inner regions of halos have weaker alignments with the 
local tidal field than the spin vectors obtained for the whole 
halos. The observational results are consistent with the alignments 
obtained for the inner regions of halos, suggesting that the disk 
material traces the angular momentum of dark matter halos 
in the inner region.

\section*{Acknowledgements}

This work is supported by the 973 Program (No. 2015CB857002), national
science foundation of China (grant Nos. 11203054, 11128306, 11121062, 11233005,
11073017), NCET-11-0879, the Strategic Priority Research Program ``The
Emergence of Cosmological Structures" of the Chinese Academy of
Sciences, Grant No. XDB09000000 and the Shanghai Committee of Science
and Technology, China (grant No. 12ZR1452800).

This work is supported by the High Performance Computing Resource
in the Core Facility for Advanced Research Computing at Shanghai
Astronomical Observatory.

\end{document}